\documentclass[preprint,prd,showpacs,amsmath,amssymb,amsthm,nofootinbib]{revtex4}
\usepackage[mathscr]{euscript}
\usepackage{bm}
\usepackage{graphicx}
\usepackage{subfigure}
\usepackage[colorlinks=true,linkcolor=red]{hyperref}
\newcommand{\bea}{\begin{eqnarray}}
\newcommand{\eea}{\end{eqnarray}}
\newcommand{\beq}{\begin{equation}}
\newcommand{\eeq}{\end{equation}}

\def\/{\over}

\begin{document}

\title{Emergent scenario in the Einstein-Cartan theory}
\author{Qihong Huang$^{1}$,  Puxun Wu$^{2,3}$  and Hongwei Yu$^{1,2,}$\footnote{Corresponding author}  }
\affiliation{$^1$ Institute of Physics and Key Laboratory of Low
Dimensional Quantum Structures and Quantum
Control of Ministry of Education,\\
Hunan Normal University, Changsha, Hunan 410081, China \\
$^2$ Center for Nonlinear Science and Department of Physics, Ningbo
University, Ningbo, Zhejiang 315211, China
\\$^3$Center for High Energy Physics, Peking University, Beijing 100080, China }

\begin{abstract}
We study the emergent scenario, which is  proposed to avoid the big bang singularity, in the Einstein-Cartan (EC) theory with a positive cosmological constant and a perfect fluid by analyzing the existence and stability of the Einstein static (ES) solutions.   We find  that there is no stable ES solution for a spatially flat or open universe. However, for a spatially closed universe,  the stable ES solution does exist, and in the same existence parameter regions, there also exists  an unstable one. With the slow decrease  of the equation of state $w$ of the perfect fluid, the stable and unstable critical points move close gradually and coincide once $w$ reaches a critical value, so that the stable critical point becomes an unstable one. As a result, if $w$ approaches a constant at $t\rightarrow -\infty$, the universe can stay at the stable ES state past eternally, and furthermore it can naturally exit from this state and evolve into an inflationary era if  $w$ decreases slowly as time goes forward. Therefore, the emergent scenario that avoids the big bang singularity  can be successfully implemented in the EC theory  of gravity.
\end{abstract}
\pacs{98.80.Cq, 04.50.Kd}

\maketitle

\section{Introduction}

One of the most basic and ancient questions in cosmology is whether the Universe has a beginning or has existed eternally, and this question has been brought into  serious discussion by using the knowledge of general relativity and modern cosmology in recent decades. However, the answer is still far from clear. The standard cosmological model implies that the Universe stems from a big bang singularity. To avoid this singularity,  Ellis $et$ $al.$\cite{Ellis1,Ellis2} proposed the so-called emergent scenario, in which the Universe stays in an Einstein static (ES) state past eternally and then evolves to a subsequent inflationary era. This scenario suggests that the Universe originates from an ES state rather than a big bang singularity. Unfortunately,  the original emergent scenario  does not seem to successfully resolve the singularity problem as expected, because
 there is no stable ES solution in general relativity~\cite{Eddington1930,Gibbons1987,Gibbons1988,Barrow2003} so that the Universe hardly stays at the initial state in a long time as a result of the existence of various perturbations.
 However, recent studies show that an implementation of the emergent scenario is possible in the modified theories of gravity. In this regard,  it  has been found that the ES state  is stable against homogeneous scalar perturbations   in various such theories including $f(R)$ gravity, $f(T)$ gravity, loop quantum gravity, and so on~\cite{Mulryne2005,Zhang2010, Wu2010, Carneiro2009, Cai2013, Bohmer2007}.

The Einstein-Cartan (EC) theory introduced by \'{E}lie Cartan in 1923 \cite{Cartan} is an interesting modified gravity, since it extends,  in a natural way,  Einstein's general relativity by including the spacetime torsion, instead of assuming it to be zero, and treating it in the same  footing as the spacetime curvature. In the EC gravity, in which the metric and the non-symmetric affine connection are two independent quantities,  the torsion arises from the matter source of spinor fields, and  it is not a dynamical  quantity~\cite{Hehl1,Hehl2,Hehl3} since there is a relation between the intrinsic angular momentum (spin) of fermionic  matter and the spacetime torsion.  Thus, one can incorporate the spinor field into the torsion-free general theory of relativity~\cite{Watanabe}.  As a result, the EC theory is  equal to the general relativity with an addition of an effective perfect fluid~\cite{Weyssenhoff,Obukhov,Berredo-Peixoto,Brechet} in the energy-momentum tensor, which is  the contribution of the spacetime torsion.    At macroscopic scales,  fermionic particles can be averaged and described as a spin fluid: the Weyssenhoff fluid~\cite{Weyssenhoff,Obukhov}. The combination of the torsion and the spin fluid  in the cosmological context of the EC theory behaves as a stiff matter with a negative energy density~\cite{Hehl4, Nurgaliev, Gasperini}, which  leads to gravitational repulsion. This  becomes very significant at extremely high densities, and it can help to avoid the curvature singularity by violating the energy condition of the singularity theorems~\cite{Trautman, Kuchowicz} and the big bang singularity through a nonsingular big bounce~\cite{Poplawski111,Ziaie}.   The EC theory has also been shown to be able  to solve the flatness and horizon problems without an exponential inflation~\cite{Poplawski1,Poplawski2,Poplawski3,Poplawski4}. Furthermore, the effects of torsion and the spinning matter in the cosmic inflation, late time acceleration, and so on,  have been studied in~\cite{Kuchowicz,Gasperini,Kim,Szydlowski,Ray,Shie,Fabbri,Vakili}.

Recently,  the existence and stability of an ES universe in the EC theory with a  Weyssenhoff perfect fluid  and a normal perfect fluid  are studied in~\cite{Atazadeh}. It has been found that there is a stable ES state in a closed spatial geometry and  the Universe can stay at this stable state  eternally. However, since this state corresponds to a center equilibrium point, our Universe cannot naturally evolve from it into an inflationary era. So it remains unclear whether an emergent scenario that avoids the big bang singularity can be successfully implemented in the EC theory of gravity and this is exactly what we are planing to address in the present paper. We systematically study the existence and  stability of the Einstein static state in all possible spatial geometries in the EC theory with a postive cosmological constant rather than  only the spatially closed case considered in \cite{Atazadeh} and in particular we demonstrate that with a positive cosmological constant the emergent scenario can be successfully realized  in the EC theory in the spatially closed universe.

The paper is organized as follows. In Section 2,  we review the cosmic equations in the framework of the EC theory. In Section 3, we analyze  the conditions of the existence of ES solutions, and discuss their stability and the phase transition from the stable  ES state to an inflation.
   Finally,  our main conclusion is presented in Section 4.  Throughout this paper, unless specified, we adopt the metric signature ($+, -, -, -$) and set $8\pi G =\kappa$. Latin  indices run from 0 to 3 and  the Einstein convention is assumed for repeated indices.

\section{The field equations of Einstein-Cartan theory}

In the EC theory of gravity, the action takes the form
\bea\label{action}
S=\int \sqrt{-g}d^4x\bigg [-\frac{1}{2\kappa}(\tilde{R}-2\Lambda)+\mathcal{L}_M\bigg]\;,
\eea
where $\tilde{R}$ is the Ricci scalar constructed by the asymmetric connection $\tilde{\Gamma}^\mu_{\; \;\alpha\beta}$ rather than the symmetric affine (Levi-Civita) connection  ${\Gamma}^\mu_{\; \;\alpha\beta}$ used in general relativity, $\Lambda$ is the cosmological constant and $\mathcal{L}_M$ is the Lagrangian density of matter.  In this theory,   the spacetime curvature and torsion have the same status with the curvature and torsion tensors being defined respectively by $\tilde{R}^\mu_{\;\;\nu\lambda\alpha}=\partial_\lambda\tilde{\Gamma}^\mu_{\; \;\nu\alpha}- \partial_\alpha\tilde{\Gamma}^\mu_{\; \;\nu\lambda}+\tilde{\Gamma}^\mu_{\; \;\beta\lambda} \tilde{\Gamma}^\beta_{\; \;\nu\alpha}-\tilde{\Gamma}^\mu_{\; \;\beta\alpha}\tilde{\Gamma}^\beta_{\; \;\nu\lambda}$  and $T^\mu_{\; \;\nu\alpha}= \tilde{\Gamma}^\mu_{\; \;\nu\alpha}-\tilde{\Gamma}^\mu_{\; \;\alpha\nu}$.  And the connection $\tilde{\Gamma}^\mu_{\; \;\alpha\beta}$ can be decomposed into two parts
\bea\label{Conn}
\tilde{\Gamma}^\mu_{\; \;\alpha\beta}=\Gamma^\mu_{\; \;\alpha\beta}+K^\mu_{\; \;\alpha\beta}\;,
\eea
where $K^\mu_{\; \;\alpha\beta}$ is the contorsion tensor,  which relates to the torsion tensor via
\bea\label{Contortion}
K^\mu_{\; \;\alpha\beta}=\frac{1}{2}(T^\mu_{\; \;\alpha\beta}-T^{\;\;\mu}_{\alpha\; \;\beta}-T^{\;\;\mu}_{\beta\; \;\alpha})\;.
\eea

Varying the action (\ref{action}) with respect to the metric tensor  and the contorsion tensor, respectively, gives the field equations of the EC theory~\cite{Hehl1,Hehl2,Hehl3}
\bea\label{FE0}
G^{\mu\nu}-\Lambda g^{\mu\nu}-(2 T_{\alpha\beta}^{\quad\beta}+\tilde{\nabla}_\alpha)(Q^{\mu\nu\alpha}-Q^{\nu\alpha\mu}-Q^{\alpha\mu\nu})=\kappa T^{\mu\nu}\;,\\ \label{FE2}
Q^{\mu\nu\alpha}=\kappa \tau^{\mu\nu\alpha}\;,
\eea
where $G^{\mu\nu}$ is the Einstein tensor, $Q_{\mu\nu}^{\quad \alpha}=T_{\mu\nu}^{\quad \alpha}+\delta^\alpha_\mu T_{\nu\beta}^{\quad \beta}-\delta^\alpha_\nu T_{\mu\beta}^{\quad \beta}$, $T^{\mu\nu}=\frac{2}{\sqrt{-g}}\frac{\delta \mathcal{L}_m}{\delta g_{\mu\nu}}$ is the energy-momentum tensor, and $\tau^{\mu\nu\alpha}=\frac{1}{\sqrt{-g}}\frac{\delta \mathcal{L}_m}{\delta K_{\mu\nu\alpha}}$ is the spin density tensor.  Eq.~(\ref{FE2}) shows that the torsion is proportional to the spin density, which indicates that the latter is the source of the former.

Combining Eqs.~(\ref{FE0}) and (\ref{FE2}), one can show that the field equations can be expressed in the standard form of general relativity with a modification in the energy-momentum tensor~\cite{Weyssenhoff, Berredo-Peixoto, Brechet} as
\bea\label{FQ}
G^{\mu\nu}- \Lambda g^{\mu\nu}=\kappa(T^{\mu\nu}+ \theta^{\mu\nu})\;,
\eea
where
\bea\label{Theta}
\theta^{\mu\nu}=-4 \tau^{\mu\alpha}_{\quad[\beta}\tau^{\nu\beta}_{\quad\alpha]}-2 \tau^{\mu\alpha\beta}\tau^{\nu}_{\;\;\alpha\beta}+\tau^{\alpha\beta\mu}\tau^{\quad \nu}_{\alpha\beta}+\frac{1}{2}g^{\mu\nu}(4\tau^{\;\;\alpha}_{\lambda\;\;\; [\beta}\tau^{\lambda\beta}_{\quad\alpha]}+\tau^{\alpha\beta\lambda}\tau_{\alpha\beta\lambda})
\eea
is the correction to the energy-momentum tensor generated by the torsion.

In the EC theory,   $T^{\mu\nu}$   can be separated as follows
\bea
T^{\mu\nu}=T_F^{\mu\nu}+T_S^{\mu\nu}\;.
\eea
Here, $T_F^{\mu\nu}=(\rho+p)u^\mu u^\nu-p g^{\mu\nu}$  represents  the usual perfect fluid   with $u_\mu$ being the four velocity,    and  $\rho$ and $p$ being the energy density and pressure, respectively. Assuming its equation of state is a constant, one has  $p=w\rho$. $T_S^{\mu\nu}$ is the energy-momentum tensor of an intrinsic-spin fluid, which reads
\bea\label{TS}
T_S^{\mu\nu}&=&u^{(\mu}S^{\nu)\alpha}u^\beta u_{\alpha;\beta}+(u^{(\mu}S^{\nu)\alpha})_{;\alpha}+T_{\alpha\beta}^{\;\;\;(\mu}u^{\nu)}S^{\beta\alpha} \nonumber \\ &&-u^\beta S^{\alpha(\nu}T^{\mu)}_{\;\;\;\alpha\beta}-\omega^{\alpha(\mu}S^{\nu)}_{\;\;\;\alpha}+u^{(\mu}S^{\nu)\alpha}\omega_{\alpha\beta}u^\beta\;,
\eea
where $\omega$ is the angular velocity associated with the intrinsic spin and a semicolon represents the covariant derivative with respect to the Levi-Civita connection.
This spin fluid can be described by the  so-called Weyssenhoff fluid, which is a continuous macroscopic medium characterized in microscopic scales by the spin of matter fields. 
The spin density of the Weyssenhoff fluid is  described by an antisymmetric tensor $S_{\mu\nu}=-S_{\nu\mu}$  and   has the form~\cite{Obukhov} \bea \tau^{\mu}_{\;\;\;\nu\alpha}=u^\mu S_{\nu\alpha}\;.
\eea

Substituting this expression into Eqs.~(\ref{Theta}, \ref{TS}), one can obtain that~\cite{Weyssenhoff, Berredo-Peixoto, Brechet}
\bea
  T_S^{\mu\nu} =-\kappa \sigma^2 u^\mu u^\nu \;,\\
  \theta^{\mu\nu} =\frac{1}{2} \kappa \sigma^2 u^\mu u^\nu +\frac{1}{4} \kappa \sigma^2 g^{\mu\nu} \;,
\eea
where \begin{equation}
\sigma^2=\frac{1}{2} S_{\mu\nu}S^{\mu\nu}
\end{equation}
is the spin density scalar.

Thus, rewriting the field equations of the EC theory as
\bea\label{Fq} G^{\mu\nu}= \kappa\tilde{T}^{\mu\nu},\eea we have
 \bea
 \tilde{T}^{\mu\nu} = T^{\mu\nu}+\theta^{\mu\nu}\ = (\rho+p-\rho_s-p_s) u^{\mu} u^{\nu}-  (p-p_s-\kappa^{-1}\Lambda)g^{\mu\nu}\;,
 \eea
 where $\rho_s=p_s=\frac{1}{4} \kappa \sigma^2$. Since $p_s/\rho_s=1$, the effect of torsion and the spin matter   can be considered as a stiff matter with negative energy density and pressure, which leads to gravitational repulsion. Here we can see that the correction to the energy-momentum tensor due to the  spin-spin interaction is of the second order in the gravitational coupling constant ($\kappa$). Thus, it is extremely weak, but it can become  very significant at extremely high energy densities in the very early universe.

 To study the ES solution in the EC theory, we consider a homogeneous and isotropic universe described by the Friedman-Lema\^{i}tre-Robertson-Walker metric:
 \bea
 ds^2=dt^2-a^2(t) \bigg[\frac{dr^2}{1-kr^2}+r^2(d\theta^2+\sin^2\theta d\phi^2)\bigg]\;,
 \eea
where $t$ is the cosmic time and $a(t)$ is the cosmic scale factor. $k=0$, $1$ or $-1$ correspond to a spatially flat, closed or open universe, respectively.  Substituting this metric into the field equations (Eq.~(\ref{Fq})),  we find that the $0-0$ component gives the Friedmann equation
 \begin{equation}\label{FE}
\dot{a}^2+k=\frac{\kappa}{3}(\rho-\rho_{s})a^2+\frac{\Lambda}{3}a^2\;,
\end{equation}
where a dot denotes the derivative with respect to the cosmic time $t$. Since the perfect fluid satisfies the continuity equation $\dot{\rho}+3\frac{\dot{a}}{a}(1+w) \rho=0$ and  $\tilde{T}^{\mu\nu}_{; \mu}=0$, 
one can obtain that~\cite{Nurgaliev,Atazadeh}
\begin{equation}\label{En}
\rho=\rho_{0}a^{-3(1+w)}, \quad \rho_{s}=\tilde{D} a^{-6}\;.
\end{equation}
Here $\rho_{0}$  and $\tilde{D}$ are two  integral constants, which are assumed to be positive.
By differentiating the Friedmann equation and using Eq.~(\ref{En}), we get the Raychadhuri equation
\begin{equation}
2\ddot{a}=-(1+3w)\frac{\dot{a}^2+k}{a}+D(1-w)\frac{1}{a^5}+\Lambda(1+w)a\;,
\end{equation}
where $D\equiv \kappa \tilde{D}$.

\section{The Einstein static solutions and their stability}
The ES solution is given by the condition $\ddot{a}=\dot{a}=0$, which implies $a=a_{Es}$ and $H(a_{Es})=0$. Thus, the Raychadhuri equation dictates  that the critical points are determined by the following cubic equation of $\frac{1}{a^2_{Es}}$
\begin{equation}\label{3E}
(1-w)D\frac{1}{a^6_{Es}}-(1+3w)\frac{k}{a^2_{Es}}+\Lambda(1+w)=0\;.
\end{equation}
And the corresponding energy density $\rho$ at the critical point satisfies
\begin{equation}\label{RE}
\kappa \rho(a_{Es})=\frac{3k}{a^2_{Es}}+\frac{D}{a^6_{Es}}-\Lambda \;,
\end{equation}
which is obtained from the Friedmann equation.  Since $a_{Es}$ is positive and $\rho(a_{Es})$ should take a nonnegative value, the existence conditions of the critical  point  are $\rho(a_{Es})\geq0$ and $a^2_{Es}>0$.

 In order to simplify our discussion, we rewrite Eqs.~(\ref{3E}, \ref{RE}) as follows:
\begin{equation}\label{3E2}
(1-w)D_{0}\frac{1}{(\Lambda a^2_{Es})^3}-(1+3w)\frac{k}{\Lambda a^2_{Es}}+(1+w)=0\;.
\end{equation}
\begin{equation}\label{3E3}
\kappa \rho(a_{Es})=\Big[\frac{3 k}{\Lambda a^2_{Es}}+\frac{D_{0}}{(\Lambda a^2_{Es})^3}-1\Big] \Lambda\;,
\end{equation}
where $D_{0}= D\Lambda^2>0$. For a cubic equation given in (\ref{3E2}),  the solutions are determined by the following expression
\begin{equation}
\Delta=B^2-4AC,
\end{equation}
where $A=-3\lambda \alpha $, $B=-9\lambda \beta$, and $C=\alpha^2 $ with $\lambda=(1-w)D_{0}$, $\alpha=-(1+3w)k$ and $\beta=(1+w)$. Respectively,
$\Delta>0$, $=0$ and $<0$ correspond to one, two, and three real solution(s).

To study the stability of critical points, we introduce two variables $x_{1}=a$ and $x_{2}=\dot{a}$, and as a result, we have
\begin{equation}
\dot{x}_{1}=x_{2},
\end{equation}
\begin{equation}
2\dot{x}_{2}=-(1+3w)\frac{x_{2}^2+k}{x_{1}}+(1-w)D\frac{1}{x^5_{1}}+\Lambda(1+w)x_{1}.
\end{equation}
The stability of these critical points are determined by the eigenvalues of the coefficient matrix,  which are obtained from linearizing the system described by the above equations near these critical points. The eigenvalue $\mu^2$ can be expressed as:
\begin{equation}\label{mu2}
\mu^2=\Big[\frac{1+3w}{2}\frac{k}{\Lambda a^2_{Es}}-\frac{5}{2}(1-w)D_{0}\frac{1}{(\Lambda a^2_{Es})^3}+\frac{1}{2}(1+w)\Big]\Lambda.
\end{equation}
If $\mu^2<0$, a small perturbation from the fixed point will result in an oscillation about this point rather than a exponential deviation. Thus,  the corresponding ES solution is a center equilibrium point. Otherwise, it is  an unstable point.

For  the case $\Lambda=0$,  from Eqs.~(\ref{3E}, \ref{RE}, \ref{mu2}), we find that
\begin{equation}
a^4_{Es}=\frac{(1-w)D}{(1+3w)k}\;, \quad \kappa \rho(a_{Es})=\frac{1}{a^2_{Es}}\frac{4k}{1-w}\;, \quad \mu^2=\frac{2D}{a^6_{Es}}(w-1)\;.
\end{equation}
So there exists a  critical point which is stable under the condition $D>0$ and $-\frac{1}{3}<w<1$ for $k=1$, as has been studied in~\cite{Atazadeh}. Apparently, the universe can stay at this ES state  eternally, but cannot exit from this state and  naturally evolve into an inflationary phase. The critical point  is unstable for $k=-1$,  although it exists under the condition $w>1$ and $D>0$.

Now, we discuss the case with a positive cosmological constant. We consider all possible spatial geometries, i.e., the spatial flat, closed and open universes rather than only the closed one with $k=1$.

\subsection{$k=0$}
For a spatially flat universe ($k=0$),   Eqs.~(\ref{3E2}) and (\ref{3E3}) reduce to
\begin{eqnarray}
\frac{D_0}{(\Lambda a^2_{Es})^3}= \frac{1+w}{-1+w}\;, \quad \kappa \rho(a_{Es})=\frac{2}{-1+w}\Lambda\;,
\end{eqnarray}
which show that the existence condition for the critical point is $w>1$.
From Eq.~(\ref{mu2}), one can see that  the eigenvalue $\mu^2=3(1+w)\Lambda$ is positive in the existence region, and this means that there is no stable ES solution in this case.

\subsection{$k=1$}

For a spatially closed  universe ($k=1$), we have  $\lambda=(1-w)D_{0}$, $\alpha =-(1+3w)$ and $\beta=(1+w)$. To analyze the solutions of Eq.~(\ref{3E2}), we need to consider the value of $\Delta$ in three different cases.

(i) $\Delta>0$: This requires $w>1$, or $w<-\frac{1}{3}$, or $-\frac{1}{3}<w<1$ with $D_{0}>-\frac{4(1+3w)^3}{27(-1+w)(1+w)^2}>0$. The only one real solution of Eq.~(\ref{3E2}) in this case, which we  label as Point $E$,  can be expressed as
\begin{equation}
Point \quad E: \quad \frac{1}{\Lambda a^2_{Es}}=-\frac{1}{3\lambda}\Big[Y^{\frac{1}{3}}_++Y^{\frac{1}{3}}_-\Big],
\end{equation}
where $Y_{\pm}=\frac{3\lambda}{2}(-B\pm\sqrt{\Delta})$. Apparently,  $\Lambda a^2_{Es}>0$ requires that $\lambda<0$ and both $Y_{+}$ and $Y_{-}$ are positive. For the case of $\lambda<0$, $w>1$ is needed, and this leads to $B>0$ and $A<0$. One can see that  $Y_{+}=\frac{3\lambda}{2}(-B+\sqrt{\Delta})<0$ since $\Delta>B^2$. This conflicts with the requirement of  $Y_{+}>0$ from $\Lambda a^2_{Es}>0$.  Thus, the critical point $E$ is physically meaningless and should be discarded.

(ii) $\Delta<0$: In this case \bea\label{Co} 0<D_{0}<-\frac{4(1+3w)^3}{27(-1+w)(1+w)^2}\,, \quad -\frac{1}{3}<w<1\eea are required, and these yield $\lambda>0$, $B<0$ and $A>0$.  Now Eq.~(\ref{3E2}) has three different real solutions and thus there exist three different critical points:
\begin{equation}
Point \quad F: \quad \frac{1}{\Lambda a^2_{Es}}=-\frac{2}{3\lambda}\sqrt{A}\cos\big(\frac{\theta}{3}\big),
\end{equation}
\begin{equation}\label{PG}
Point \quad G: \quad \frac{1}{\Lambda a^2_{Es}}=\frac{\sqrt{A}}{3\lambda}\Big[\cos\big(\frac{\theta}{3}\big)+\sqrt{3}\sin\big(\frac{\theta}{3}\big)\Big],
\end{equation}
\begin{equation}
Point \quad H: \quad \frac{1}{\Lambda a^2_{Es}}=\frac{\sqrt{A}}{3\lambda}\Big[\cos\big(\frac{\theta}{3}\big)-\sqrt{3}\sin\big(\frac{\theta}{3}\big)\Big],
\end{equation}
where $\theta=\arccos(T)$ with $T=-\frac{3}{2}\lambda BA^{-\frac{3}{2}}$. It is easy to see that  $0<T<1$, and as a result,  $0<\theta<\frac{\pi}{2}$.
Since $\lambda>0$,  $\Lambda a^2_{Es}<0$ for point $F$. Thus, this point is not physically meaningful.

In the region $0<\theta<\frac{\pi}{2}$,  $\cos\big(\frac{\theta}{3}\big)>\sqrt{3}\sin\big(\frac{\theta}{3}\big)>0$, and consequently $\Lambda a^2_{Es}$ for critical  points $G$ and $H$ are positive. In addition, we find that if conditions given in (\ref{Co}) are obeyed, $\rho(a_{Es})$ is also positive for these two critical points. Thus, both points $G$ and $H$ are  physical under the conditions given in  (\ref{Co}).

(iii) $\Delta=0$: This implies that $D_{0}=-\frac{4(1+3w)^3}{27(-1+w)(1+w)^2}$ with $-\frac{1}{3}<w<1$. In this case, there are two different real solutions. We find that critical points $G$ and $H$ coincide because of $\theta=0$. Thus, two different critical points are  point $F$ with $\theta=0$ and point $G$ or $H$ with $\theta=0$. As in the case of $\Delta<0$,  point $F$ is still physically meaningless since it implies $\Lambda a^2_{Es}<0$, and as a result, only point $G$ with $\theta=0$  needs to be considered.

\begin{figure}[!htb]
                \centering
                \includegraphics[width=0.457\textwidth ]{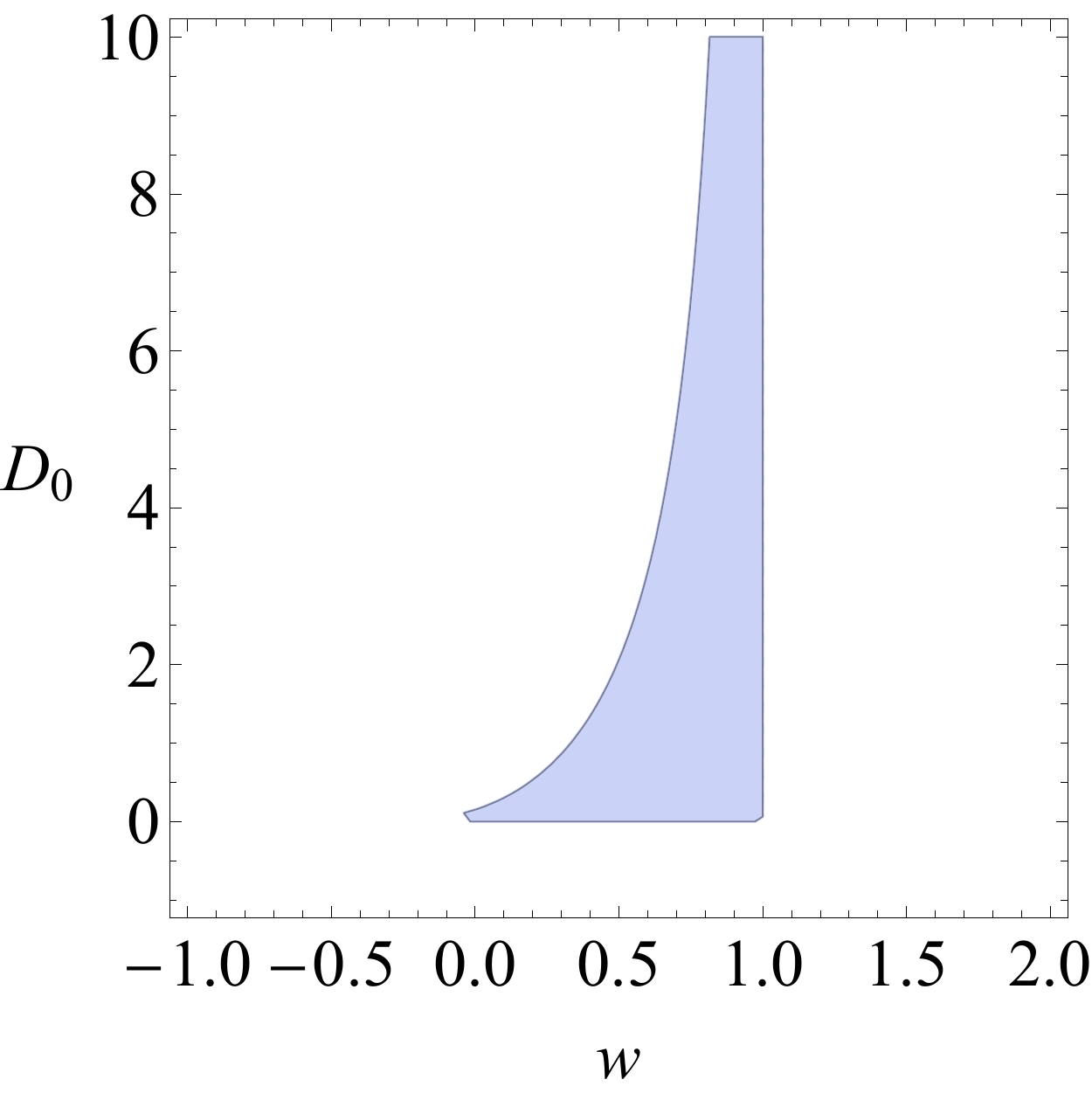}
                \caption{\label{Fig1} Regions of stability for point $G$ in the $(w,D_{0})$ parameter space.}
  \end{figure}

 Now, we discuss the stability of  critical points $G$ and $H$ by analyzing their eigenvalues shown in Eq.~(\ref{mu2}). We find by numerical calculations that only point $G$ is always stable as long as it exists, and the regions of stability for Point $G$ in the $(w,D_{0})$ parameter space are shown in Fig.~(\ref{Fig1}).  Point $H$ is always unstable. A summary of the existence and stability of all critical points is given in Tab.~(\ref{Tab1}).

\begin{table}
\caption{\label{Tab1} Summary of the existence and stability of critical points for $k=1$.}
\begin{center}
 \begin{tabular}{|r|r|r|r|}
  \hline
  \hline
  &Critical point&Existence&Stability\\
  \hline
  $\Delta>0$ & Point $E$ & no & $-$ \\
  \hline
  $\Delta<0$ & Point $F$ & no & $-$ \\
  & Point $G$ & yes & stable if it exists \\
  & Point $H$ & yes & unstable \\ \hline
  $\Delta=0$ & Point $F$ with $\theta=0$ & no & $-$ \\
  & Point $G$ or $H$ with $\theta=0$ & yes & unstable \\
 \hline
 \hline
  \end{tabular}
\end{center}
   \end{table}

\begin{figure}[!htb]
                \centering
                \includegraphics[width=0.457\textwidth ]{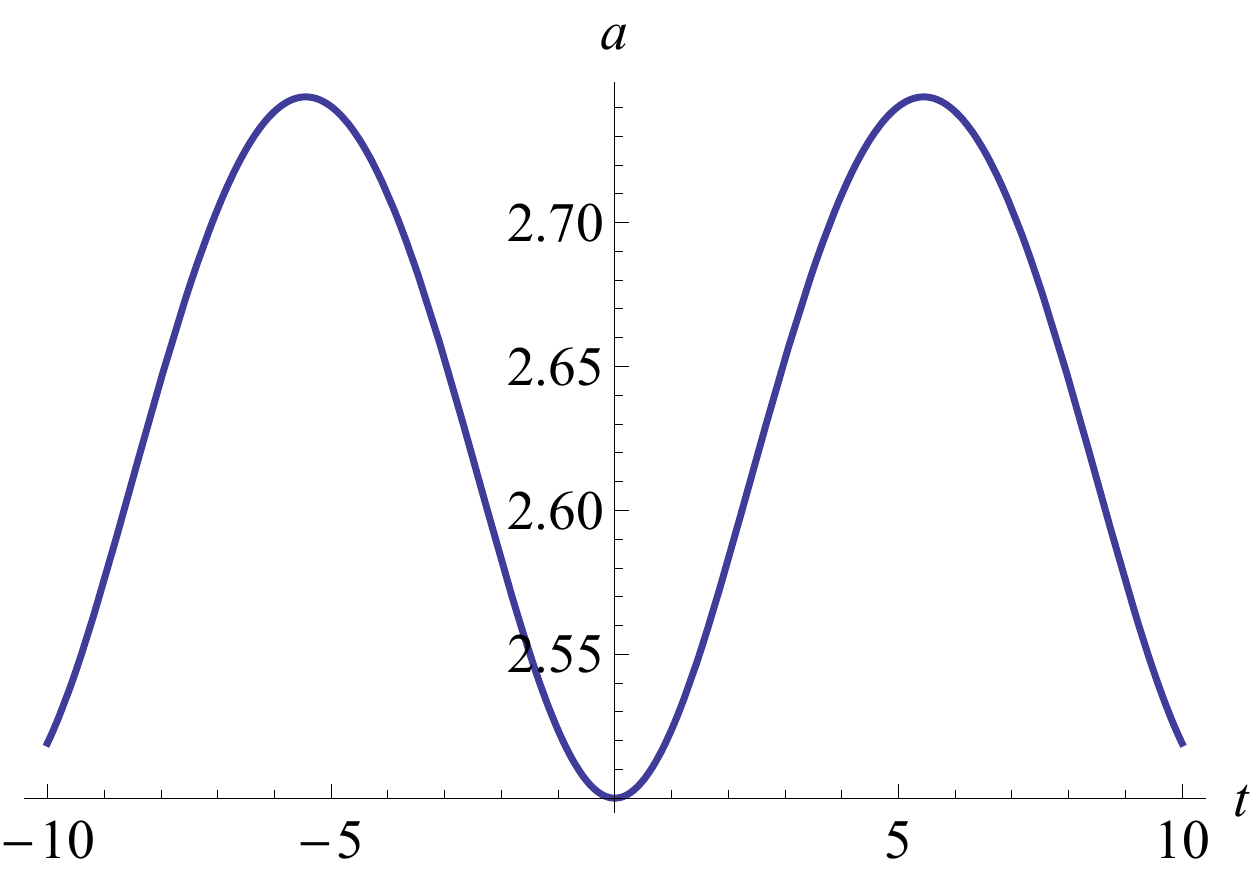}
                \includegraphics[width=0.457\textwidth ]{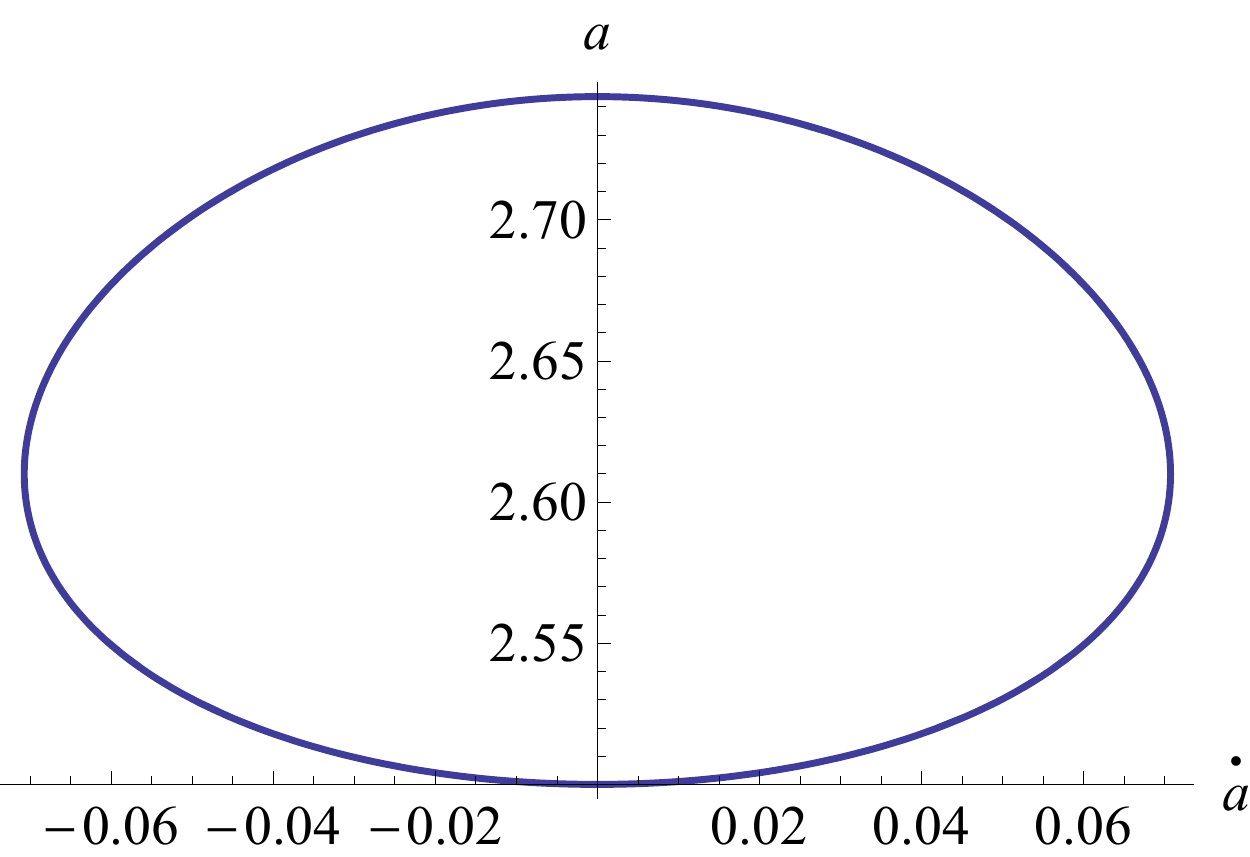}
                \caption{\label{Fig2} Evolutionary curve of the scale factor with time and the corresponding phase diagram in space $(a,\dot{a})$ with $w=0.6$, $\Lambda=0.1$ and $D_{0}=2$.}
        \end{figure}

\begin{figure}[!htb]
                \centering
                \includegraphics[width=0.47\textwidth ]{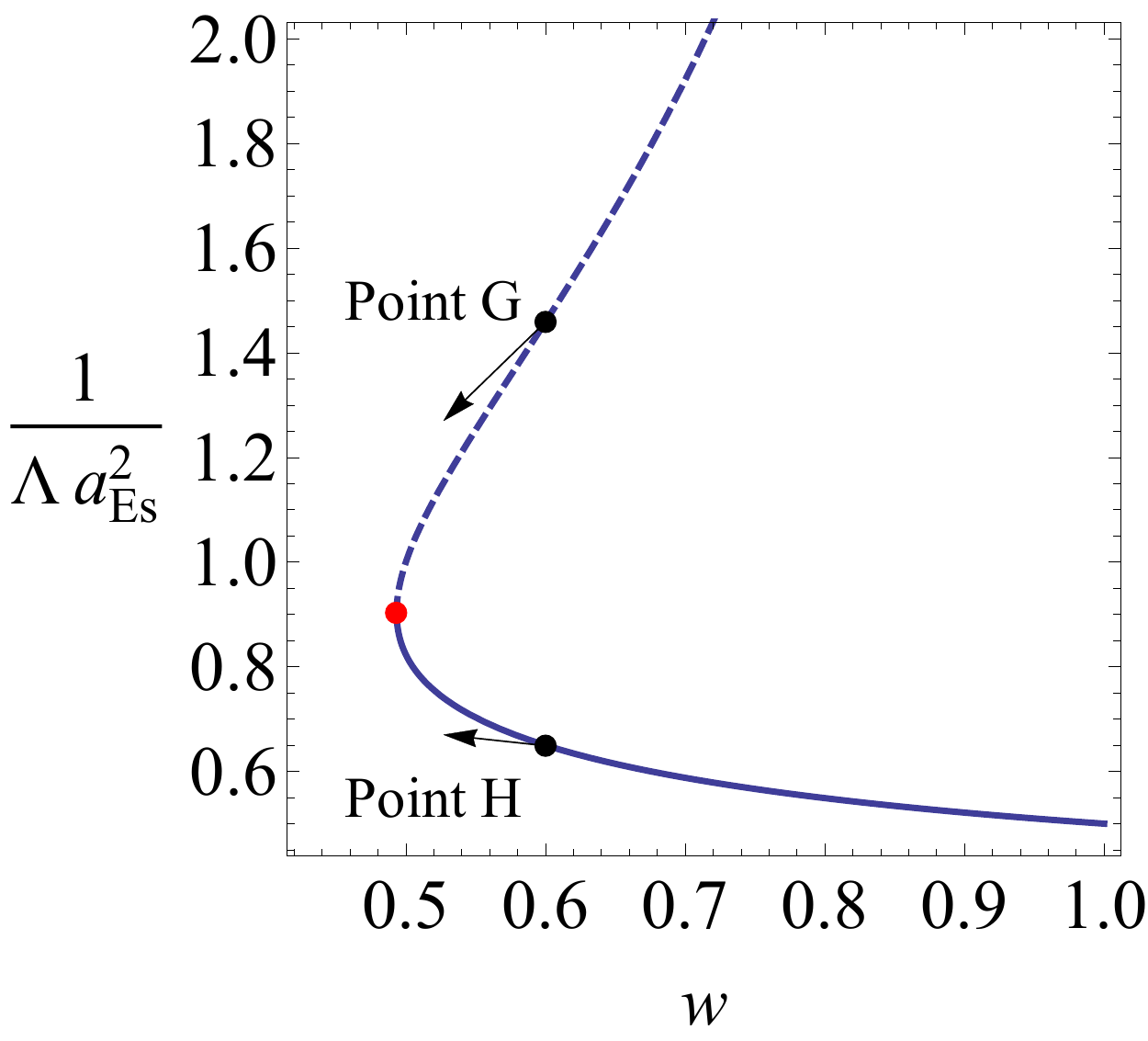}
                \includegraphics[width=0.43\textwidth ]{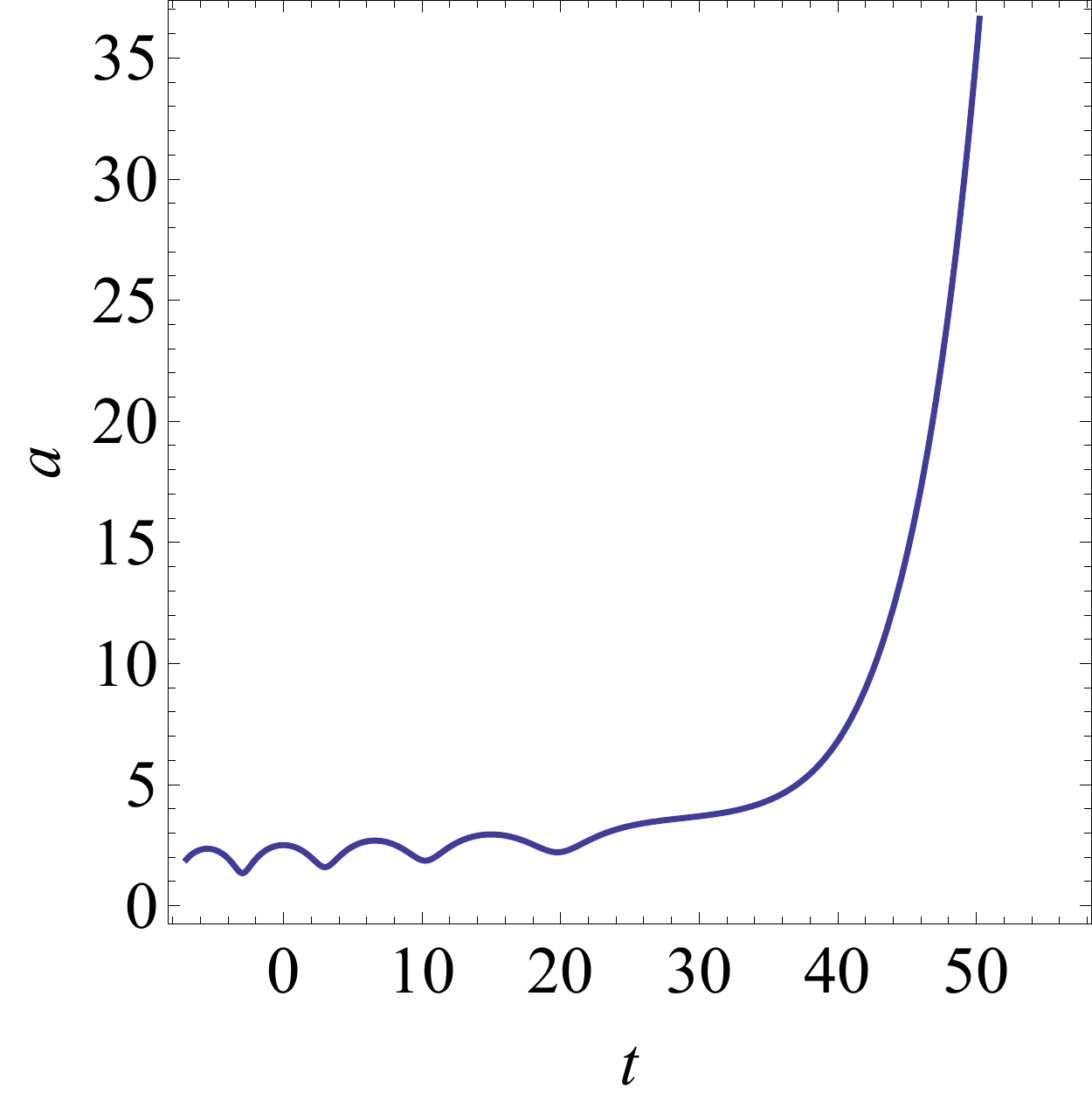}
                \caption{\label{Fig3} The evolutions of the stable point $G$ and unstable point $H$ with $D_{0}=2$ and the decrease of $w$ are shown in the left panel, while the right panel shows the phase transition from a stable state to an inflationary one with $D_{0}=2$,  $\Lambda=0.1$ and a slowly decreasing  $w$. } 
        \end{figure}

Thus, if our Universe stays at point $G$ initially, it can stay at this state past eternally since it is a stable center point. 
Now we want to make an estimate about the size of the Universe at this initial stable state. For this purpose, let us note that, with Eq.~(\ref{PG}), the  scale factor in this stable state can be written as 
\bea\label{aEs}
a_{Es}= \bigg({D} \frac{3(1-w)}{3w+1}\bigg)^{1/4}\Big[\cos\big(\frac{\theta}{3}\big)+\sqrt{3}\sin\big(\frac{\theta}{3}\big)\Big]^{-1/2}\; .
\eea 
To estimate the value of $a_{Es}$, we take $D_0=1$ and $w=0.6$, which are  in stable region shown in Fig.~(\ref{Fig1}), and then we get
\bea
a_{Es}\simeq \frac{1}{\sqrt{\Lambda}}\;.
\eea  
Assuming $\rho_\Lambda=\kappa^{-1}\Lambda$ to be the inflation  energy scale ($\sim 10^{90} \, {\rm g/cm^3}$), one has $a_{Es}\sim 10^{-31} {\rm cm}$, which is larger than the Planck length by about two orders of magnitude.
If the initial value of the cosmic scale factor deviates slightly from the value given by Eq.~(\ref{aEs}), the Universe will undergo an infinite oscillation around the center equilibrium point, as shown in Fig.~(\ref{Fig2}).  
However, a successful emergent scenario that avoids the big bang singularity demands not only a stable ES state that the Universe can stay past eternally but also a natural exit from it into an inflationary era.  Next we show that this is possible. For this purpose,  we assume that the equation of state of the perfect fluid approaches  a constant at $t\rightarrow-\infty$ and decreases very slowly with time going forward.  With the decrease of $w$, the stable critical point $G$ and unstable point $H$ will move close gradually and coincide once $w$ reaches a critical value in which $\Delta=0$.  So, the stable point $G$ becomes an unstable one.   In other words, if the cosmic scale factor satisfies the initial condition which is given in Eq.~(\ref{aEs}), our Universe can evolve from a stable state to an unstable one with the decrease of $w$, as shown in the left panel of  Fig.~(\ref{Fig3}). As a result, the Universe can naturally exit from the stable state and enter an inflationary phase. In  the right panel of Fig.~(\ref{Fig3})  we  show this phase transition. Finally, let us note that  our of analysis of the emergent scenario does not take the quantum effects of gravity into consideration which presumably come into play at very high energies in the very early universe. Fortunately, in the emergent scenario we just described  here,  the Universe starts from 
an initial stable state which is about two orders of magnitude larger  than the Planck length, so the effects from the quantization of the gravitational field may not  be significant and  can be safely neglected.

\subsection{$k=-1$}

In this case, one has  $\lambda=(1-w)D_{0}$, $\alpha =(1+3w)$ and $\beta=(1+w)$.

(i) $\Delta>0$: This leads to \begin{eqnarray}
D_{0}>\frac{4(1+3w)^3}{27(-1+w)(1+w)^2}>0\;,\quad  |w|>1\;,\\ D_{0}>\frac{4(1+3w)^3}{27(-1+w)(1+w)^2}>0\;,\quad -1<w<-\frac{1}{3}\;,  \\ or \quad  \quad    -\frac{1}{3}<w<1\;. \end{eqnarray}  There is one real solution,  labeled as Point $I$, which has the same form as Point $E$. As have been discussed in the case of $k=1$, $\Lambda a^2_{Es}>0$ requires that $\lambda<0$ and both $Y_{+}$ and $Y_{-}$ are positive. Since $\lambda <0$ means $w>1$,  which gives $B>0$ and $A>0$, both $Y_{+}$ and $Y_{-}$ are positive.  We find that if the conditions  \begin{eqnarray}\label{PI}
w>1\;,\quad  D_{0}>\frac{4(1+3w)^3}{27(-1+w)(1+w)^2}>0 \end{eqnarray} are satisfied,   $\rho(a_{Es})$ is also positive. Thus, Eq.~(\ref{PI}) gives the existence condition for Point $I$.

(ii) $\Delta<0$: This  inequality leads to  $0<D_{0}<\frac{4(1+3w)^3}{27(-1+w)(1+w)^2}$ with $|w|>1$ or $-1<w<-\frac{1}{3}$. Eq.~(\ref{3E2})  has three real solutions in this case.  We name them as Points $J$, $K$ and $L$,  which have the same forms as Points $F$, $G$ and $H$.   One can obtain that $A>0$ and $-1<T<1$, which imply $0<\theta<\pi$, $\cos(\frac{\theta}{3})>0$ and $\sin(\frac{\theta}{3})>0$.

For Point $J$, $\Lambda a^2_{Es}>0$ requires $\lambda<0$. If $0<D_{0}<\frac{4(1+3w)^3}{27(-1+w)(1+w)^2}$ and $w>1$  are satisfied,  both $\Lambda a^2_{Es}$  and $\rho(a_{Es})$ are positive. Thus, different from Point $F$, this critical point is physically meaningful.

For Point $K$, the $\Lambda a^2_{Es}>0$ condition requires $0<D_{0}<\frac{4(1+3w)^3}{27(-1+w)(1+w)^2}$ with $-1<w<-\frac{1}{3}$ or $w<-1$, while the one for Point $L$ demands $0<D_{0}<\frac{4(1+3w)^3}{27(-1+w)(1+w)^2}$ with $-1<w<-\frac{1}{3}$. $\rho(a_{Es})$ for Points $K$ and $L$
 are negative when $\Lambda a^2_{Es}>0$. Thus, these two critical points are not  physical.

(iii) $\Delta=0$: This means that  $D_{0}=\frac{4(1+3w)^3}{27(-1+w)(1+w)^2}$ with $-1<w<-\frac{1}{3}$ or $|w|>1$. There are two different real solutions which are Point $J$ with $\theta=0$ and Point $K$ or $L$ with $\theta=0$. As in the case of $\Delta<0$, Point $J$ is physical under the condition $w>1$ because of $\Lambda a^2_{Es}=-\frac{2\sqrt{A}}{3\lambda}>0$ and $\kappa \rho(a_{Es})=\frac{2(7+9w)\Lambda}{(-1+w)(1+3w)}>0$. For Point $K$ or $L$ with $\theta=0$, the requirements from $\Lambda a^2_{Es}>0$  and $\rho(a_{Es})>0$  conflict with each other and  so this critical point is not physical.

In summary, in the case of $k=-1$, Point $I$, Point $J$ and Point $J$ with $\theta=0$ are all physical, but none of them is stable. Thus, there is no stable ES solution in this case.

\section{Conclusion}

The EC theory is a modification of Einstein's general  relativistic theory of gravitation by introducing spacetime torsion. However, the spacetime torsion is not a dynamical quantity and it can be expressed completely in terms of the spin sources. The EC theory of gravity can be expressed as  the general relativity theory with an exotic stiff perfect fluid. Recently, the existence and stability of ES universe in the EC theory with a positive spatial curvature have been discussed in~\cite{Atazadeh}, and it has been found that a stable ES solution exists so that the universe can stay at the stable state  eternally. This suggests that the EC theory supports a non-singular cosmology.  However, this cosmology is not realistic though non-singular, since the universe can not exit from the stable state  and  naturally evolve into an inflationary era.

In this paper, we reanalyze the existence and stability of ES solutions in the EC theory by adding a positive cosmological constant. In addition to the case of a positive spatial curvature, we also study the cases of $k=0$ and $-1$. We find  that in spatially flat and open universes there is no stable ES solution. While, for the spatially closed universe ($k=1$), the stable ES solution exists, and in the same parameter regions  there also exists  an  unstable one. With the slow decrease of the equation of state $w$ of the perfect fluid, the stable and unstable critical points move close gradually. Once $w$ reaches a critical value, as shown in the left panel of Fig.~(\ref{Fig3}), the stable critical point coincides with the unstable one, and becomes an unstable point. Thus, if $w$ is a constant at $t\rightarrow -\infty$, the Universe can stay at the stable ES state past eternally. Then, it can  naturally  evolve into an inflationary era with a slowly decreasing $w$. Therefore, in the EC theory,  a successful emergent scenario to avoid the big bang singularity can be successfully implemented.

\begin{acknowledgments}
This work was supported by the National Natural Science Foundation of China under Grants No. 11175093, No. 11222545, No. 11435006, and No. 11375092;  the  Specialized Research Fund for the Doctoral Program of Higher Education under Grant No. 20124306110001.

\end{acknowledgments}

\end{document}